\documentclass[12pt]{iopart}

\usepackage{bm,graphicx,textcomp,amssymb,dcolumn,float,color,comment,bbold,cancel}
\usepackage{tikz}
\usepackage{pgfplots}[compat=1.18]
\pgfplotsset{compat = newest}

\newcommand{\ket}[1]{\left|#1\right>}
\newcommand{\bra}[1]{\left<#1\right|}

\newcommand{\nn}{\nonumber\\}

\newcommand{\bea}{\begin{eqnarray}}
\newcommand{\ea}{\end{eqnarray}}
\newcommand{\eea}{\end{eqnarray}}
\newcommand{\ord}{\,{\cal O}}
\newcommand{\li}{\,\widehat{\cal L}}

\def\h{\hat}
\def\s{{}^\dagger}

\usepackage{romannum}

\usepackage{braket,mleftright}

\begin{document}

\title{Quasi-particle propagation across semiconductor--Mott insulator interfaces}
\author{Jan Verlage$^1$, Friedemann Queisser$^{2,3}$, Nikodem Szpak$^1$, Jürgen König$^1$, Peter Kratzer$^1$ and Ralf Schützhold$^{2,3}$}
\address{$^1$Fakult\"at f\"ur Physik and CENIDE, Universit\"at Duisburg-Essen,
  Lotharstra{\ss}e 1, 47057 Duisburg, Germany}
  \address{$^2$Helmholtz-Zentrum Dresden-Rossendorf, 
Bautzner Landstra{\ss}e 400, 01328 Dresden, Germany}
\address{$^3$Institut f\"ur Theoretische Physik, 
Technische Universit\"at Dresden, 01062 Dresden, Germany}

\ead{jan.verlage@uni-due.de}

\begin{abstract}
As a prototypical example for a heterostructure combining a weakly and a 
strongly interacting quantum many-body system, we study the interface 
between a semiconductor and a Mott insulator.
Via the hierarchy of correlations, we derive and match the propagating 
or evanescent (quasi) particle solutions on both sides
{\color{black} and assume that the interactions among the electrons 
in the semiconducting regions can be absorbed by an effective potential.}
While the propagation is described by a band-like dispersion in both   
the weakly and the strongly interacting case, 
the inverse decay length across the interface follows a different dependence on the band gap in the Mott insulator and the semiconductor. 
As one consequence, tunnelling through a Mott insulating layer behaves quite 
differently from a semiconducting (or band insulating) layer. 
For example, we find a strong suppression of tunnelling for energies in the 
middle between the upper and lower Hubbard band of the Mott insulator. 
\end{abstract}
\maketitle

\section{Introduction}

The proper understanding of strongly interacting quantum many-body systems is one 
of the grand challenges of theoretical physics. 
Weakly interacting systems which can be treated via perturbation theory with 
respect to the small coupling strength typically display universal features. 
In contrast, strongly interacting systems feature new phenomena and require the 
development of different approaches and methods, such as 
strong-coupling perturbation theory \cite{pairault2000strong,iskin2009strong,senechal2002cluster,sherman2015mott},
dynamical mean-field theory \cite{georges1996dynamical,tsuji2014nonequilibrium,peters2014spin}, 
numerical diagonalization \cite{PhysRevLett.98.180601,PhysRevLett.105.250401,PhysRevA.79.021608,PhysRevLett.101.063001} {\color{black}or numerical methods like Quantum Monte Carlo \cite{golor2015nonlocal,zujev2013induced}}.

The treatment of a heterostructure combining a weakly and a strongly interacting 
quantum many-body system will then be even more challenging because the approach 
suited for describing the strongly interacting system 
(e.g., strong-coupling perturbation theory) may not be applicable to the weakly 
interacting system and vice versa. 
Moreover, a related situation occurs in electronic transport through a system
where the electronic interaction in one subsystem is strong, e.g.,
through a molecular layer or quantum dots. The need to describe such systems
has prompted the development of a special branch of density functional theory,
the so-called i-DFT for stationary currents \cite{stefanucci2015,jacob2020}.
In the present work, we employ a discrete lattice model of solids that provides
a mathematical description of systems with a well-defined amount of electronic
correlations.
As a prototypical example for such a heterostructure, we study the interface 
between a semiconductor.
%
{\color{black} To simplify our analysis, we will assume that the interactions among the electrons within the semiconducting region can be represented by an effective potential. 
This approach aligns with Fermi-liquid theory, where electron-electron interactions are incorporated by renormalizing parameters of the free electron gas \cite{solovyev2017renormalized,landau1957theory}.
}

The fabrication of these systems \cite{doi:10.1126/sciadv.aaz0611,noh2014direct}
as well as the theoretical investigation how properties carry over through the interfaces \cite{yonemitsu2007suppression,jiang2012density,zhao2010mott,helmes2008kondo} {\color{black}and the interfaces themselves \cite{okamoto2005interface,okamoto2006lattice}} has advanced over the years. {\color{black}Numerical methods like the determinant quantum Monte
Carlo (DQMC) method are used to investigate magnetic properties of a heterostructure of alternating metallic and Mott
insulating layers \cite{zujev2013induced,jiang2012density}.}

{\color{black} An alternative to studying heterostructures in the lateral direction is the usage of stacked heterostructures bound by van der Waals forces \cite{castellanos2022van}. These stacked structures offer a versatile 'playground' for exploring systems that combine different material classes, enabling the integration of various components such as single-band Mott insulators \cite{PhysRevX.13.041049}, the Mott insulating state of FeSe \cite{kang2024mott}, or layered Mott insulating materials like FePX3 (X = S, Se) \cite{jin2022mott}, alongside semiconducting van der Waals materials. Although these stacked heterostructures also contain interfaces between different regions, and could thus be described by the same formalism we use here, our discussion and wording will focus specifically on the case of lateral structures.
}

The goal is to derive the propagating or evanescent (quasi) particle modes and
to match them across the interface.
To this end, we employ the method of the hierarchy of correlations which is based 
on a formal expansion into powers of the coordination number $Z$ 
(which is assumed to be large) \cite{PhysRevA.82.063603,queisser2019environment,PhysRevA.89.033616_Queisser}.
In contrast to many other methods, this approach can be applied to the regions 
of weak and strong interactions on the same footing and allows us to derive
the propagation of (quasi) particle modes.
As a major result, we give explicit expressions for the  transmitted and
reflected current densities as functions of energy. Remarkably, the tunnelling current through a Mott insulating layer vanishes in the middle of the Mott band gap due to destructive interference of the particle and hole channel.

\section{Extended Fermi-Hubbard Model}

In order to start with a tight-binding model which is capable of describing the relevant (valence or conduction) band  of a 
semiconductor as well as a Mott insulator, we employ the extended Fermi-Hubbard 
model \cite{hubbard1963electron}
\bea
\label{Fermi-Hubbard}
\hat H=-\frac1Z\sum_{\mu\nu s} T_{\mu\nu} \hat c_{\mu s}^\dagger \hat c_{\nu s} 
+\sum_\mu U_\mu \hat n_\mu^\uparrow\hat n_\mu^\downarrow
+\sum_{\mu s}V_\mu\hat n_{\mu s}  
\,.\;
\ea
As usual, $\hat c_{\mu s}^\dagger$ and $\hat c_{\nu s}$ denote the fermionic 
creation and annihilation operators at the lattice sites $\mu$ and $\nu$ 
with spin $s\in\{\uparrow,\downarrow\}$ and $\hat n_{\mu s}$ are the 
associated number operators. 
The hopping matrix $T_{\mu\nu}$ equals the tunneling strength $T$ for nearest 
neighbors $\mu$ and $\nu$ and is zero otherwise. 
Here, we assume that the Mott insulator and the semiconductor have the same 
hopping strength $T$, but our results can be generalized to different 
hopping strengths $T$ for the two regions or for the different directions 
in a straightforward manner. 
The coordination number $Z$ counts the number of nearest neighbors $\mu$
for a given lattice site $\nu$ and is assumed to be large, $Z\gg1$.
The first term in the Hamiltonian is the kinetic energy. The second term describes the Coulomb interaction of the charged fermions at the same lattice site; an occupation of two fermions at the same site comes with an energy penalty $U$. The last term describes an additional on-site potential.
Later on, we will consider a half-space filled with a Mott insulator and a half-space filled by a semiconductor adjacent to each other. For this geometry, we want to investigate the reflection and transmission of (quasi) particles.

The on-site repulsion $U_\mu$ and potential $V_\mu$ are used to distinguish 
the Mott insulating from the semiconducting region. 
For vanishing $U_\mu$ (i.e., within the semiconductor),
the above Hamiltonian is bi-linear and can thus be diagonalized easily. This yields a bandstructure $E=V-T_\mathbf{k}$. 
Shifting the on-site potential $V$ shifts the semiconductor band egde.
In this work, our focus lies on the Mott insulator (with non-zero $U_\mu$)
and its interface to the semiconductor.

{\color{black}Heterostructures involving Mott insulating regions have been studied using various mean-field approaches. Notably, the Hartree-Fock approximation has been used to describe the electronic structure within an effective single-electron framework \cite{okamoto2004electronic,okamoto2004theory}. 
Additionally, transport through a Mott insulating region \cite{okamoto2007nonequilibrium} and charge order at the interface \cite{okamoto2004spatial,helmes2008kondo} has been examined using Dynamical Mean-Field Theory (DMFT), which approximates individual Mott layers as infinite-dimensional Mott-Hubbard systems. In contrast, we will take a different approach, employing the hierarchy of correlations method, which allows for a controlled expansion of correlation functions in large dimensions.
}
\subsection{Hierarchy of correlations}

The hierarchy of correlations was first introduced in \cite{PhysRevA.82.063603}; its basic idea is as follows: We consider the reduced density matrices, $\hat\rho_\mu$ of one lattice site, 
$\hat\rho_{\mu\nu}$ of two, and so on. Moreover, we split up the correlated parts 
via $\hat\rho_{\mu\nu}^{\rm corr}=\hat\rho_{\mu\nu}-\hat\rho_{\mu}\hat\rho_{\nu}$, 
and similarly for three-site and higher-order correlations. 
Based on the large coordination number assumption $Z\gg1$, we may employ an expansion into powers 
of $1/Z$ where we find that higher-order correlators are successively
suppressed. 
The two-point correlator scales as $\hat\rho_{\mu\nu}^{\rm corr}=\ord(1/Z)$,
while the three-point correlation is suppressed as 
$\hat\rho_{\mu\nu\lambda}^{\rm corr}=\ord(1/Z^2)$, and so on.  

Mathematically, the hierarchy starts from the von-Neumann equation ($\hbar=1$)
\begin{equation}
    i \partial_t \h \rho = \big[H,\h \rho \big] \, .
\end{equation}
We trace out all but one (two, three,...) lattice site(s) to deduct the evolution equations for the reduced density matrix
\bea
      i \partial_t \h \rho_\mu &=& \tr_{\cancel{\mu}}\big[H,\h \rho \big], \\
       i \partial_t \h \rho_{\mu\nu} &=& \tr_{\cancel{\mu}\cancel{\nu}}\big[H,\h \rho \big],
\eea
and so on.
Separating the reduced density matrices into a (trivial) product of lower-order and a correlated part, i.e. $\hat\rho_{\mu\nu}^{\rm corr}=\hat\rho_{\mu\nu}-\hat\rho_{\mu}\hat\rho_{\nu}$ for the two point correlations, we can express the evolution equations formally in the way 
\bea
\label{evolution}
i\partial_t \hat\rho_\mu 
&=& 
F_1(\hat\rho_\mu,\hat\rho_{\mu\nu}^{\rm corr})
\,,\nn
i\partial_t \hat\rho_{\mu\nu}^{\rm corr} 
&=& 
F_2(\hat\rho_\mu,\hat\rho_{\mu\nu}^{\rm corr},\hat\rho_{\mu\nu\lambda}^{\rm corr})
\, ,
\ea
i.e., apart from the single-site occupation, only the correlated parts enter.
In addition, the (yet to be determined) functions $F_1$, $F_2, \ldots$ depend on the Hamiltonian.  Generally this method works for all lattice Hamiltonians that can be written as \\
$\h H = \sum_a \h H_a + \frac{1}{Z} \sum_{a,b}\h H_{a,b}$
.
We may now approximate them by using the expansion into powers of $1/Z$. 
The first evolution equation then becomes 
$i\partial_t \hat\rho_\mu = F_1(\hat\rho_\mu,0)+\ord(1/Z)$.  
Its zeroth-order solution $\hat\rho_\mu^0$ yields the mean-field background. 
For a more detailed view see \ref{app_sec_hierarchy} or \cite{PhysRevA.82.063603,PhysRevA.89.033616_Queisser}.

In order to describe the Mott insulator, we can choose from two possibilities, the first one being a homogeneous state at half-filling,
\bea
\label{mean-field}
\hat\rho_\mu^0
=
\frac{\ket{\uparrow}_\mu\!\bra{\uparrow}+\ket{\downarrow}_\mu\!\bra{\downarrow}}{2}
\,,
\ea
{\color{black}that is spin-unpolarized.} This might be achieved by frustration or by a large enough temperature to destroy the spin ordering. As the spin modes energy $\sim T^2/U$ is much smaller than the Mott gap $U$ of the charge modes, this temperature still could be cold for the latter while destroying spin ordering.
The semiconductor is represented by 
$\hat\rho_\mu^0=\ket{\uparrow\downarrow}_\mu\!\bra{\uparrow\downarrow}$
for the valence band and by 
$\hat\rho_\mu^0=\ket{0}_\mu\!\bra{0}$ for the conduction band
(at zero temperature). 
As the Coulomb interaction in the Hamiltonian Eq.~\ref{Fermi-Hubbard} is only intra, and not inter, band it does not contribute here in the semiconductor; but for a state with a small electron or hole density it needed to be considered, see Section~\ref{Outlook}.
Note that the two cases are related to each other via particle-hole duality
(such that it would be enough to consider one case only). 

The propagation of (quasi) particles on top of the background $\h \rho_\mu^0$ is then governed by the linearization
\bea
	i \partial_t \h\rho^{\rm corr}_{\mu\nu} &\approx& 
F_2(\hat\rho^0_\mu,\hat\rho_{\mu\nu}^{\rm corr},0) \nn
 &=&
 \big[\h H_\mu ,\h\rho^{\rm corr}_{\mu\nu}] + \frac{1}{Z}[\h H_{\mu\nu},\h\rho^0_\mu\h\rho^0_\nu ]
 -
 \frac{\h\rho^0_\mu}{Z}\tr_\mu\left( [\h H_{\mu\nu}+H_{\nu\mu},\h\rho^0_\mu\h\rho^0_\nu]\right)\nn
	&&+
 \frac{1}{Z}\sum_{\alpha \neq\mu\nu}\tr_\alpha\left([\h H_{\mu\alpha}+\h H_{\alpha \mu},\h\rho^{\rm corr}_{\mu\nu}\h\rho^0_\alpha+\h\rho^{\rm corr}_{\nu\alpha}\h\rho^0_\mu]\right)\nn
&&+
 (\mu\leftrightarrow\nu)+\mathcal{O}\left(1/Z^2\right).
\ea
of the second equation~\ref{evolution}. 

\subsection{Quasi-particle operators}

%
To analyze the evolution of the (quasi) particles, it is useful to introduce the (quasi) particle and
hole operators 
\bea
\hat c_{\mu s I}=\hat c_{\mu s}\hat n_{\mu\bar s}^I=
\left\{
\begin{array}{ccc}
 \hat c_{\mu s}(1-\hat n_{\mu\bar s}) & {\rm for} & I=0 
 \\ 
 \hat c_{\mu s}\hat n_{\mu\bar s} & {\rm for} & I=1
\end{array}
\right.
\,,
\ea
where $\bar s$ denotes the spin state opposite to $s$. 
For example 
$\hat c_{\mu\uparrow I}=\ket{0}_\mu\!\bra{\uparrow}$ 
creates an empty 
site $\mu$ for $I=0$ while
$\hat c_{\mu\uparrow I}=\ket{\downarrow}_\mu\!\bra{\uparrow\downarrow}$
annihilates a doubly occupied lattice site $\mu$ for $I=1$.
For the semiconductor, these operators directly correspond to the
electrons and holes.
In the Mott insulator, however, they
%
%
are approximately, but not exactly equal to the
quasi-particle creation and annihilation operators for holons and doublons, 
see, e.g. \cite{PhysRevResearch.2.022046}. {\color{black} The usage of such effective operators that better describe the physical excitations in strongly correlated materials dates back to Hubbard X operators \cite{hubbard1965electron, ovchinnikov2004hubbard} and composite operators \cite{mancini2004hubbard, avella20021d}. }
Their time evolution is given by the von-Neumann equation, resulting in
\bea 
i \partial_t \h c_{\mu s I}&=&  -\frac{1}{Z}\sum_{\kappa \neq \mu} T_{\mu \kappa} \h c_{\kappa s I}+ \left(V_\mu +U^I_\mu \right) \h c_{\mu s I} \nn&&-\frac{1}{Z}\sum_{\kappa \neq \mu} T_{\mu \kappa} \left( \h c\s_{\kappa \bar s} c_{\mu \bar s} c_{\mu s}- c_{\mu s} c\s_{\mu \bar s} c_{\kappa \bar s} \right).
\ea

In terms of those operators $\h c_{\mu s I}$, we find to first order the following evolution equations for the 
correlation functions (for $\mu\neq\nu$)
\bea
\label{corr-evolution}
i\partial_t
\langle\hat c^\dagger_{\mu s I}\hat c_{\nu s J}\rangle^{\rm corr}
=
\frac1Z\sum_{\lambda L} T_{\mu\lambda}
\langle\hat n_{\mu\bar s}^I\rangle^0
\langle\hat c^\dagger_{\lambda s L}\hat c_{\nu s J}\rangle^{\rm corr}
\nn
-
\frac1Z\sum_{\lambda L} T_{\nu\lambda}
\langle\hat n_{\nu\bar s}^J\rangle^0
\langle\hat c^\dagger_{\mu s I}\hat c_{\lambda s L}\rangle^{\rm corr}
\nn
+
\left(U_\nu^J-U_\mu^I+V_\nu-V_\mu\right) 
\langle\hat c^\dagger_{\mu s I}\hat c_{\nu s J}\rangle^{\rm corr}
\nn
+\frac{T_{\mu\nu}}{Z}
\left(
\langle\hat n_{\mu\bar s}^I\rangle^0
\langle\hat n_{\nu s}^1\hat n_{\nu\bar s}^J\rangle^0
-
\langle\hat n_{\nu\bar s}^J\rangle^0
\langle\hat n_{\mu s}^1\hat n_{\mu\bar s}^I\rangle^0
\right) +\mathcal{O}(1/Z^2)
\,.
\ea
Here we used the abbreviation $U_\mu^I=IU_\mu$, i.e., $U_\mu^I=U_\mu$ for 
$I=1$ and $U_\mu^I=0$ for $I=0$.
Quite intuitively, the repulsion is only felt by the particle contributions 
$I=1$ but not by the hole contributions $I=0$. Expectation values denoted as $\langle \rangle^0$ are to be taken in the zeroth order solution $\h \rho^0_\mu$.

The stationary source term in last line of Eq.~\ref{corr-evolution}
can be used to derive the ground-state
correlations.
However, it is not relevant for (quasi) particle propagation which corresponds
to solutions 
$\langle\hat c^\dagger_{\mu s I}\hat c_{\nu s J}\rangle^{\rm corr}$ 
oscillating in time. 
These oscillating solutions can be derived from the first three lines 
(i.e., the homogeneous part) of the above equation. Moreover, we see that the propagation in different spin sectors is decoupled.

\subsection{Factorization}
Before describing the propagation of quasi-particles through the interface, we first need to define them for the homogeneous systems. In this case, by neglecting the source term in Eq.~\ref{corr-evolution},   
it can be shown that these oscillating solutions
can be obtained by the following factorization ansatz \cite{navez2014quasi,PhysRevA.89.033616_Queisser}
\bea
\langle\hat c^\dagger_{\mu s I}\hat c_{\nu s J}\rangle^{\rm corr}
=p_{\mu s I}^* p_{\nu s J}
\,,
\ea
where the factors $p_{\nu s I}$ obey the simple equation 
\bea
\left(i\partial_t-U^I_\mu-V_\mu\right)p_{\mu s I}=
\frac{-1}{Z}\sum_{\nu J} T_{\mu\nu} \langle\hat n_{\mu\bar s}^I\rangle^0 p_{\nu s J}
\,.
\ea
This now reduces the four equations for $\langle\hat c^\dagger_{\mu s I}\hat c_{\nu s J}\rangle^{\rm corr}$ with $I,J \in \{0,1\}$ down to two for $p_{\mu s I}$ with $I \in \{0,1\}$, while still describing the same physics as the full set of correlation functions.
To find the (quasi) particle modes, we now seek oscillating solutions with
eigen-energies $E$, i.e., $i\partial_t\to E$.

\subsection{Comparison to other approaches {\color{black} and scaling in real materials}}

Since our approach is based on an expansion into powers of $1/Z$ instead of a
ratio of coupling constants such as $U/T$ (ordinary perturbation theory)
or $T/U$ (strong-coupling perturbation theory), we may directly apply it to
both, the semiconducting and the Mott insulating region, which are then
treated on the same footing.
The quasi-particle dispersion relation within the Mott insulator can also be
studied via other approaches such as the Hubbard approximation \cite{corr_ins_julich,hubbard1964electron}, but the
$1/Z$ expansion employed here does also facilitate a systematic way to
incorporate higher orders, see Section~\ref{Outlook} below and \cite{PhysRevA.100.053617,queisser2019boltzmann}. 
The life-time broadening of bands that turn out to be important for describing (inverse) photo-emission experiments~\cite{Hoepfner13,Hansmann13} appears in higher orders of $1/Z$.

Furthermore, in contrast to methods which are based on the mapping onto an
effective single lattice site, such as dynamical mean-field theory (DMFT)~\cite{Kotliar96}
or its time-dependent version (t-DMFT), our approach is capable of taking
the space-time dependence fully into account, especially in more than one
spatial dimension \cite{queisser2023hierarchy,krutitsky2014propagation}.
{\color{black} The DMFT assumption of $Z\to \infty$ fixes the effective bandwidth of $T/\sqrt{Z}$ \cite{RevModPhys.68.13,PhysRevB.45.6479}. Signatures of this can also be seen in the hierarchy of correlations with the $1/Z$ scaling when studying quench dynamics in finite dimension \cite{PhysRevB.109.195140}. For a fixed effective bandwidth, the {\color{black} asymptotic } magnitude of double occupations becomes independent of the dimension.} 
{\color{black} This suggests that the hierarchy of correlations is stable.}
{\color{black}Moreover, in \cite{PhysRevB.109.195140} the influence of higher order effects (and hence the scaling) is investigated for cubic lattices in several dimensions.}

{\color{black}The hierarchy works at finite $Z$. In contrast to DMFT with the $1/\sqrt{Z}$ scaling, where only the non-trivial limit of $Z\to \infty$ is well understood, the limit of infinite coordination number in the hierarchy of correlation`s $1/Z$ scaling yields a trivial result; all lattice sites decouple.}
{\color{black} Finally, the resulting equations are simple enough to allow for an
analytic treatment while DMFT is primarily a numerical method.}

{\color{black}
In this work, we assume a hypercubic lattice for simplicity, but this choice is also relevant for real materials. Many Mott insulators, particularly perovskites, exhibit Mott insulating behavior on three-dimensional cubic lattices with 
$Z=6$ nearest neighbors, as seen in materials like TiF3 \cite{sheets2023mott} and LaTiO3 \cite{maznichenko2024fragile}. Additionally, there are Mott insulators on (quasi-)two-dimensional triangular lattices with $Z=6$ neighbors, such as those reported in \cite{PhysRevB.94.161105, tomeno2020triangular, PhysRevLett.99.256403}. Importantly, since our hierarchy of correlations is governed by the coordination number rather than the dimensionality, the scaling applies equally well to both types of Mott insulators in real materials. Furthermore, second-order effects can be incorporated via a renormalized hopping parameter \cite{PhysRevA.89.033616_Queisser}, ensuring that the effects we predict remain valid across these systems.} {\color{black} Right now, these higher order calculations only exist for homogeneous systems and are subject of future works.}

\section{Mott-semiconductor interface}

To simplify the analysis, we consider a planar interface which coincides 
with a lattice plane. 
Then assuming a highly symmetric lattice such as a hyper-cubic lattice, 
we apply a Fourier transformation of the dependence parallel to the interface 

\bea
p_{\mu s I }&=\frac{1}{\sqrt{N^\parallel}}\sum_{\mathbf{k}^\parallel}p_{n,\mathbf{k}^\parallel,s I}e^{i  \mathbf{k}^\parallel \cdot\mathbf{x}_\mu^\parallel}\\
T_{\mu\nu}&=\frac{Z}{N^\parallel}\sum_{\mathbf{k}^\parallel} T_{m,n,\mathbf{k}^\parallel}e^{i \mathbf{k}^\parallel \cdot\left(\mathbf{x}_\mu^\parallel-\mathbf{x}_\nu^\parallel\right)} \,.
\eea
which yields, the parallel hopping contribution $T_{\bf k}^\|$ 
(for details, consult Appendix B). 

For better readability, we drop the $n$, $s$, and ${\bf k}^\|$ indices 
from now on, $p_{n,\mathbf{k}^\parallel,s I}\equiv p_\mu^I$, and arrive at 
\bea
\label{difference}
\left(E-U_\mu^I-V_\mu\right)p_\mu^I
+\langle\hat n_{\mu}^I\rangle^0
\sum_J T_{\bf k}^\| p_\mu^J
=
\nn
-T\frac{\langle\hat n_{\mu}^I\rangle^0}{Z}
\sum_J \left(p_{\mu-1}^J+p_{\mu+1}^J\right) 
\,,
\ea
where $\mu\in\mathbb Z$ now just labels the lattice sites perpendicular 
to the interface.  

The above equations show that the particle $p_\mu^1$ and hole $p_\mu^0$ 
contributions are linearly dependent.
In the semiconductor where one of the two expectation values 
$\langle\hat n_{\mu}^I\rangle^0$ is always zero, we find that 
either $p_\mu^0$ or $p_\mu^1$ vanishes (conduction or valence band).
In the Mott insulator, on the other hand, where these expectation values
read $\langle\hat n_{\mu}^I\rangle^0=1/2$ {\color{black} for the spin-unpolarized background}, particle $p_\mu^1$ and hole 
$p_\mu^0$ contributions are both present. 

In those regions where $U_\mu$, $V_\mu$, and $\langle\hat n_{\mu}^I\rangle^0$
are constant, we may solve the difference equation~\ref{difference} by the 
following ansatz 
\bea
\label{ansatz} 
p_\mu^I=\alpha^I e^{i\kappa\mu}+\beta^I e^{-i\kappa\mu}
\,.
\ea
Insertion of this ansatz into Eq.~\ref{difference} can then be used to 
determine the effective wavenumber $\kappa$ perpendicular to the interface.  

\subsection{Semiconductor}
\label{subseq:semicon}

In order to describe the semiconductor, we choose a vanishing repulsion 
$U_\mu=0$ and a constant on-site potential $V_\mu=V$. $\hat\rho_\mu^0=\ket{\uparrow\downarrow}_\mu\!\bra{\uparrow\downarrow}$
describes the valence band and  
$\hat\rho_\mu^0=\ket{0}_\mu\!\bra{0}$ the conduction band
(at zero temperature). 
Hence,
either one of $\langle \h n_\mu^I \rangle^0$ is zero. Consequently, it follows the simple equation
\bea
\left(E-V\right)p_\mu^I
+
 T_{\bf k}^\| p_\mu^J=
\frac{-T}{Z}
 \left(p_{\mu-1}^I+p_{\mu+1}^I\right) 
\,.
\ea
This is solved by the effective wavenumber
\bea
\label{semiconductor}
\cos\kappa_\mathrm{semi}=\frac{Z}{2T}\left[V-E-T_{\bf k}^\|\right]
\,.
\ea
Solving this equality for $E$ reproduces the well-known dispersion relation 
$E=V-T_{\bf k}$. 
As usual, real $\kappa$ describe propagating solutions while imaginary 
$\kappa$ correspond to evanescent modes, for example in tunneling.  
Note that $|\kappa|$ grows for increasing $V$ in this case, such that the 
probability for tunneling through a layer of finite thickness decreases. 
\subsection{Mott insulator}

The Mott insulator state corresponds to a large on-site repulsion $U_\mu=U\gg T$ 
which we assume to be a constant. 
In addition, we assume a vanishing potential $V_\mu=0$. 
In this case, the particle and hole solutions are related via $(E-U)p_\mu^1=E p_\mu^0$, and the effective wavenumber of their propagation reads 

\bea
\cos\kappa_\mathrm{Mott}=\frac{Z}{2T}\left[\frac{E(U-E)}{E-U/2}-T_{\bf k}^\|\right]
\,.
\label{kappaMott}
\ea
In the strongly interacting regime $U\gg T$ considered here, 
real solutions for $\kappa_\mathrm{Mott}$ exist only for small $E$ (lower Hubbard band) 
or for $E\approx U$ (upper Hubbard band). 
Again, this is consistent with the dispersion relation 
\bea
\label{Mott-dispersion}
E_{\pm}=\frac12\left(U-T_{\bf k}\pm\sqrt{U^2+T_{\bf k}^2}\right) 
\, ,
\ea
for quasi-particles and holes, respectively.  
{\color{black} The dispersion relations derived from various approximation methods reveal a similar structure {\color{black}\cite{avella1998hubbard,hubbard1963electron,herrmann1997magnetism,roth1969electron,beenen1995superconductivity}} that coincides with the one found here by our hierarchy of correlations method. The Hubbard\,\Romannum{1} approximation \cite{hubbard1963electron} offers a basic two-pole structure with distinct lower and upper Hubbard bands, capturing the Mott insulator behavior but lacking momentum dependence. Roth’s two-pole approximation enhances this by incorporating momentum-dependent terms \cite{roth1969electron,beenen1995superconductivity}, resulting in a more realistic description at intermediate coupling strengths. In contrast, the Second-Order Decoupling Approximation (SDA) \cite{herrmann1997magnetism} adopts a mean-field-like approach, yielding a narrower bandgap and less precise spectral features. Lastly, Composite Operator Methods (COM) \cite{avella1998hubbard, avella20021d, mancini2004hubbard} provide a detailed framework that accounts for complex excitations and self-energy corrections, leading to a richer dispersion relation while also allowing for higher-pole approximations \cite{avella2014hubbard} to refine the pole structure of the dispersion relation.} {\color{black} The mentioned methods rely on the truncation of an infinite series of expectation values or correlation functions (see, e.g., the discussion after eq.\,56 in \cite{roth1969electron}) which makes it a priori uncontrolled, while the hierarchy of correlation is a controlled expansion into a small control parameter $1/Z$. In addition, the methods work best for thermal states but not necessarily for dynamics which we are interested in (for thermal states within the hierarchy see \cite{queisser2023hierarchy}).}

Already at this stage, we find an important qualitative difference to a 
semiconductor.
According to Eq.~\ref{semiconductor}, $|\kappa_\mathrm{semi}|$ grows for increasing $V$ 
as $|\kappa_\mathrm{semi}|\sim\ln V$. 
In contrast, $|\kappa_\mathrm{Mott}|$ remains finite in the strongly interacting limit 
$U\to\infty$ of the Mott insulator.
On the other hand, $|\kappa_\mathrm{Mott}|$ diverges for $E=U/2$, i.e., exactly in the middle 
between the upper and lower Hubbard bands, resulting in a vanishing tunneling probability.

\subsection{Reflection and transmission amplitudes}

Now we are in the position to study the propagation of (quasi)
particles across the interface.
Starting from the Mott insulator region, we may employ the following ansatz  

\bea
\left(
\begin{array}{c}
p_\mu^0 \\
p_\mu^1
\end{array}
\right) 
=
{\cal N} 
\left(
\begin{array}{c}
E-U \\
E 
\end{array}
\right) 
\left[e^{i\kappa_\mathrm{Mott}\mu} + Re^{-i\kappa_\mathrm{Mott}\mu}\right] 
\, ,
\ea
where ${\cal N}$ is a normalization constant.
As expected, the hole contributions $p_\mu^0$ dominate for the lower Hubbard band 
(small $E$) whereas the particle contributions $p_\mu^1$ dominate for the upper 
Hubbard band ($E\approx U$).
In the semiconductor region, on the other hand, we have exactly $p_\mu^0=0$ 
for the conduction band (where $\hat\rho_\mu^0=\ket{0}_\mu\!\bra{0}$) 
or  $p_\mu^1=0$ for the valence band (where 
$\hat\rho_\mu^0=\ket{\uparrow\downarrow}_\mu\!\bra{\uparrow\downarrow}$). 

Matching these two contributions across the interface
allows us to derive reflection and transmission amplitudes, see \ref{app:reflection_am} for the spin-unpolarized one. Based on  
Eq.~\ref{difference}, we obtain 
the reflection amplitude 
\bea
\label{eq:refl-amplitude}
R=-\frac{1-\exp\{-i(\kappa_{\rm M}-\kappa_{\rm semi})\}}
{1-\exp\{i(\kappa_{\rm M}+\kappa_{\rm semi})\}}
\,.
\ea
where $\kappa_{\rm semi}$ and $\kappa_{M}$ are solutions of Eq.~\ref{semiconductor} and \ref{kappaMott}, respectively.
Analogous to the case of impedance matching in electrodynamics, for example, 
we find $R=0$ for $\kappa_{\rm M}=\kappa_{\rm semi}$.

In the continuum limit of small $\kappa$, we recover the standard expression 
$R=(\kappa_{\rm M}-\kappa_{\rm semi})/(\kappa_{\rm M}+\kappa_{\rm semi})$
which yields prefect reflection $R\to1$ for strong impedance mismatch 
$\kappa_{\rm M}\gg\kappa_{\rm semi}$ or
$\kappa_{\rm M}\ll\kappa_{\rm semi}$. 

In a similar way, the transmission amplitude is obtained as 
\bea 
\label{eq:trans-amplitude}
\mathcal{T}=\mathcal{N}(2E-U)(1+R) \,.
\ea

To allow for a probability interpretation, we need to define a probability current. From the (quasi) particle evolution equation Eq.~\ref{difference} we deduct 
\begin{equation}
\label{eq:current_hier}
    \partial_t |p_\mu^0+p_\mu^1|^2+j_\mu-j_{\mu-1}=
iU\left[p_\mu^0 \left(p_\mu^1\right)^*-p_\mu^1 \left(p_\mu^0\right)^*\right]
\end{equation}
with 
\bea
j_\mu&=i\,\frac{T}{Z}\left(p_\mu^0+p_\mu^1\right)\left(p_{\mu+1}^0+p_{\mu+1}^1\right)^*+c.c. \,.
\ea
For the present case of a single mode, the ratio $p^1/p^0$ is real, so the right side of Eq.~\ref{eq:current_hier} is zero. With this, we get a continuity equation which
 permits the interpretation of $j_\mu$ as a probability current.

\begin{figure}[t!]
    \centering
   	\begin{tikzpicture}
		\begin{axis}[
			xmin = -0.338516, xmax = 1.2,
			ymin = 0, ymax = 1.0,
			xtick distance = .2,
			ytick distance = 0.2,
			minor tick num = 1,
			major grid style = {lightgray},
			minor grid style = {lightgray!25},
			width = \linewidth,
			height = 0.7\linewidth,
			xlabel = {$E/U$},
			ylabel = {$\frac{j^\mathrm{trans}}{j^\mathrm{in}}$},
			legend cell align = {left},
            legend style={at={(0.7,1)}}
			]

            \addplot[smooth,
			thin,
			] table[x index=0, y index=1] {pics/interface_tab_schar.dat};

            \addplot[smooth,
			thin, orange, loosely dashed
			] table[x index=0, y index=2] {pics/interface_tab_schar.dat};

            \addplot[smooth,
			thin, blue, dashed
			] table[x index=0, y index=3] {pics/interface_tab_schar.dat};

            \addplot[smooth,
			thin, red, dashdotted
			] table[x index=0, y index=4] {pics/interface_tab_schar.dat};
			
			\legend{$V=-0.1U$,$V=0.2U$,$V=1U$,$V=1.3U$}

			\draw[dotted] (axis cs:0.838516,0) -- (axis cs:0.838516,6);
				\draw[dotted] (axis cs:0,0) -- (axis cs:0,6);		
    \draw[dotted] (axis cs:1,0) -- (axis cs:1,6);		
            \draw[dotted] (axis cs:-0.238516,0) -- (axis cs:-0.238516,6);
    
		\end{axis}
	
	\end{tikzpicture}
    \caption{(Color online) Transmission probability through the Mott insulator semiconductor interface as function of the energy $E$ for different on-site potentials $V$ in a two-dimensional lattice with $T_\mathbf{k}^\|=\frac{2T}{Z}$ and $T=0.4U$. The gray dotted lines depict the positions of the lower (from $-0.25E/U$ to $0E/U$) and upper ($0.85E/U-1E/U$) Hubbard band of the Mott side. The semiconductor band shifts with $V$.}
    \label{fig:trans_interface}
\end{figure}
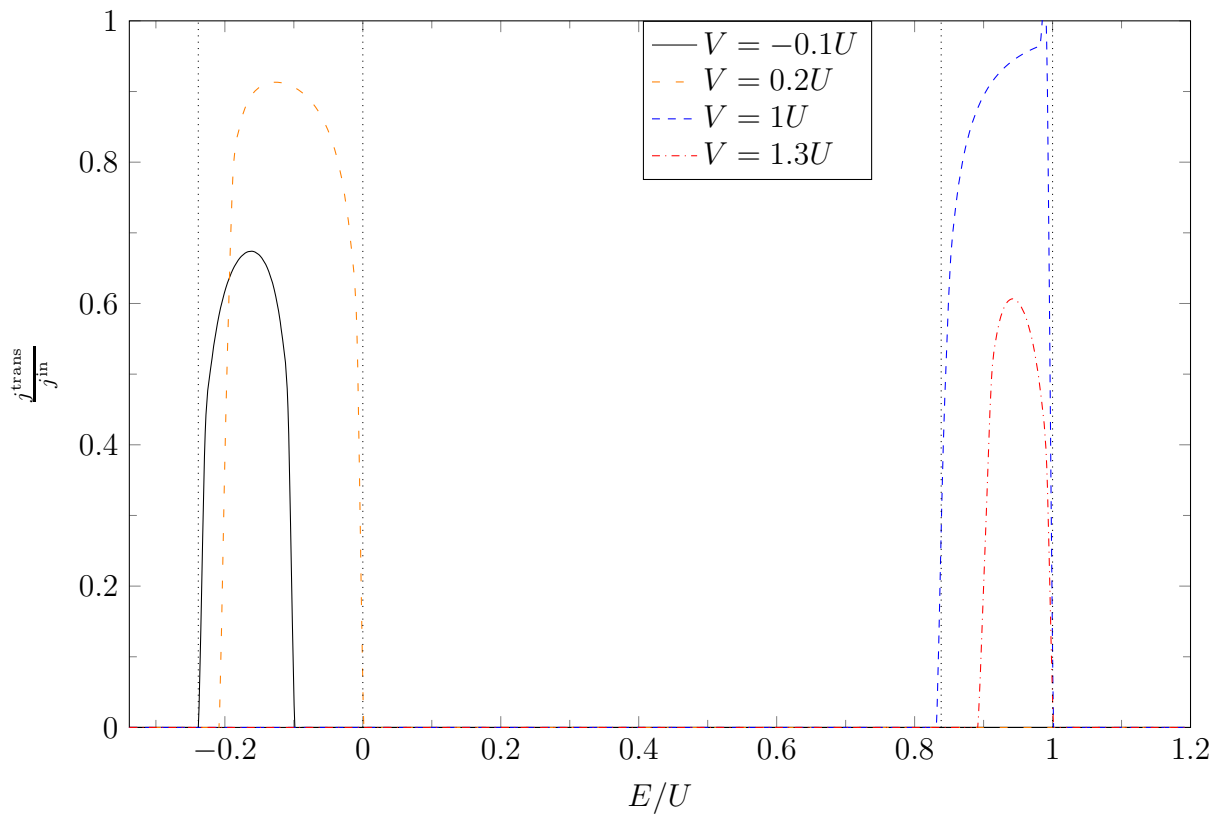

The transmission probability $j^\mathrm{trans}/j^\mathrm{in}$ shows similar features as known from the transmission between a semiconductor and a conventional charge-transfer insulator, see e.g. \cite{im2009tunneling,doi:10.1021/acsomega.9b04011,koberidze2018structural}. This is shown in Fig.~\ref{fig:trans_interface}: 
only if the semiconductor band edge falls into the lower or upper Hubbard band, shown between gray-dotted lines in Fig.~\ref{fig:trans_interface},
appreciable transmission is found. The parameter $V$ is used to tune the semiconductor band edge through the relevant range of energies. 
However, in contrast to a conventional insulator, the transmission across the interface into the Mott insulator drops exactly to zero (in the approximation used),  if $V$ falls right in between the upper and lower Hubbard band. 

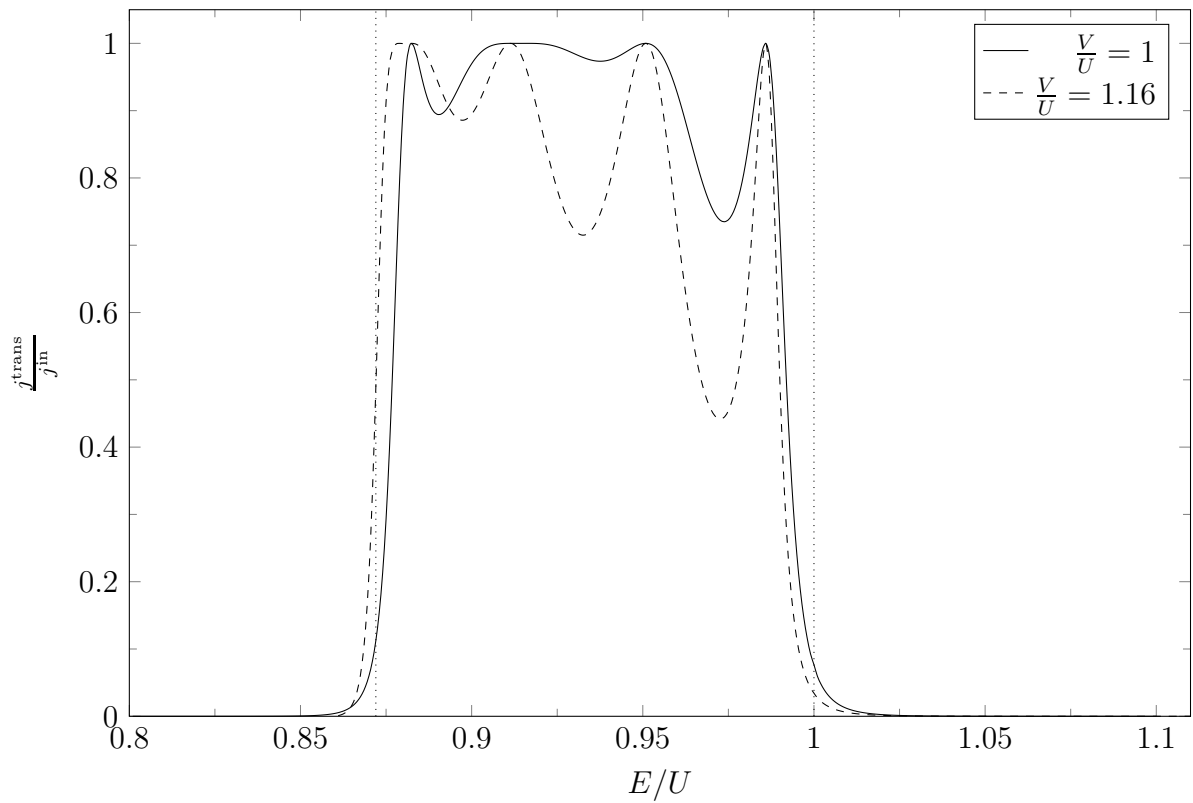
\begin{figure}[t!]
    \centering
   	\begin{tikzpicture}
		\begin{axis}[
			xmin = 0.8, xmax = 1.11,
			ymin = 0, ymax = 1.05,
			xtick distance = .05,
			ytick distance = 0.2,
			minor tick num = 1,
			major grid style = {lightgray},
			minor grid style = {lightgray!25},
			width = \linewidth,
			height = 0.7\linewidth,
			xlabel = {$E/U$},
			ylabel = {$\frac{j^\mathrm{trans}}{j^\mathrm{in}}$},
			legend cell align = {right},
			]

			\addplot[
			smooth,
			thin,
			] file[skip first] {pics/layer.dat};

            \addplot[
			smooth,
			thin,
            dashed   
			] file[skip first] {pics/layer116.dat};
			
			\legend{$\frac{V}{U}=1$,$\frac{V}{U}=1.16$}

			\draw[dotted] (axis cs:0.872015,0) -- (axis cs:0.872015,6);
				\draw[dotted] (axis cs:1,0) -- (axis cs:1,6);			
		\end{axis}

	\end{tikzpicture}
    \caption{Transmission through the upper Hubbard band for $T=0.3U$ in a two-dimensional lattice with $T_\mathbf{k}^\|=\frac{2T}{Z}$ for two different semiconductor potentials $V=U$ and $V=1.16U$. The dotted vertical lines mark the upper Hubbard band, and the conduction bands of the semiconductor are given approximately by the plot range. Above the Hubbard band and below, so between the two bands, there is a tunnelling current. Inside the band there are peaks to perfect transmission for energies that fulfil the resonance condition $\kappa_\mathrm{Mott}\cdot d = z \pi$ with $z \in \mathbb{Z}$ with the barrier width $d$, they appear for all $V$ at the same position.}
    \label{fig:trans_layer}
\end{figure}

The method presented here can be extended to the case of a finite number of $d$ Mott-insulator layers sandwiched between two semiconductor leads. 
The Mott layer acts like a potential barrier for the electrons, similar to the results known from single-particle physics. 
If the energy $E \ge U$, a channel for propagation of a (quasi)
particle through the barrier opens up, and transmission resonances are observed as function of the 
energy. This is illustrated for typical parameter values in Fig.~\ref{fig:trans_layer}.
For $0< E < U$ tunneling through the barrier is observed. 
 For a large number of layers $d \gg 1$ and $e^{- \kappa_{\rm M}}<1$ the transmission in this tunneling regime is given by 
\bea
\frac{j_n^\mathrm{trans}}{j_n^\mathrm{in}}\approx
4 e^{- 2d\kappa_\mathrm{M} }(1-e^{-2\kappa_\mathrm{M}})^2(1-\cos(\kappa_\mathrm{semi}))^2
\ea

with
\bea
e^{- \kappa_\mathrm{Mott}} \approx \frac{4T(E-U/2)}{Z U^2}
\ea
in the strong coupling regime (derivation in \ref{app:Mott_layer}).
{\color{black}
The tunneling through a conventional insulator would exhibit a suppression of the tunneling current that depends exponentially on the spatial width of the insulating region. In contrast, for a Mott insulator, the tunneling current vanishes precisely at the center of the Mott insulating gap, i.e. $E=U/2$,  regardless of the width of the Mott insulating region.}

In \ref{app:toy_model} we present a toy model with few Mott sites between semiconducting leads to analyse and better understand the reason for this vanishing of the transmission. 
It is found that there is a minimum of the transmission resulting from a coherent superposition of transmission amplitudes involving the lower and the upper Hubbard band. The destructive interference of these two transmission amplitudes is responsible for the suppression of the tunneling.

\section{Conclusions}

In summary, we studied the interface between a weakly interacting semiconductor
and a strongly interacting Mott insulator.
To this end, we carried out the formal expansion into inverse powers of the
coordination number $1/Z$ establishing a hierarchy of correlations.
This method, in contrast to others like strong coupling perturbation theory
or perturbation theory, is suited both for the strongly and weakly interacting
systems.
Moreover, it works for general lattices $T_{\mu \nu}$ and is particularly suited
for higher dimensions.
Using this method we employed the extended Fermi-Hubbard model to describe
the semiconductor with a homogeneous on-site potential $V$ and the Mott
insulator with the on-site Coulomb repulsion $U$.

We established effective particle and hole operators $\h c_{\mu sI}$ on a
half-filled mean-field background and used them to calculate the time evolution
of the resulting correlation functions.
The homogeneous part of these equations give rise to propagating
(quasi) particle solutions.
These solutions recover the well known dispersion relations for the
semiconductor $E=V- T_\mathbf{k}$ as well as the two Hubbard bands at
$E \approx 0$ and $E\approx U$.
We calculate an analytical expression for the reflected and transmitted
(quasi) particle current at the interface and see that this is analogous
to the case of impedance matching in electrodynamics.
Lastly, already from the dispersion relation of the Mott insulator {\color{black} in the spin-unpolarized state} we deduct
that tunneling is strongly suppressed
if the energy is right in the middle between the upper and lower Hubbard bands
due to destructive interference of the tunneling amplitudes from particle and
hole contributions. 
This is confirmed 
by the analytical expression for the tunneling current through a
semiconductor-Mott-semiconductor heterostructure.

\section{Outlook}\label{Outlook}

As already mentioned, the $1/Z$ expansion employed here facilitates a
systematic approach for including higher orders in $1/Z$.
For example, by inserting the solution for the two-point correlations
$\hat\rho_{\mu\nu}^{\rm corr}$ back into the first line of Eq.~\ref{evolution},
we can derive the back-reaction of these fluctuations onto the mean field.
Since the spatial translation invariance is broken by the interface,
this could generate a space charge layer -- especially at finite temperature,
which can also be taken into account within the $1/Z$ expansion \cite{queisser2023hierarchy}.

Going into another direction, one could incorporate higher-order correlations,
for example spin correlations.
In the Mott state, effective Heisenberg type spin-spin interactions due to
virtual hopping processes tend to generate anti-ferromagnetic order.
They can alter the quasi-particle dispersion relation \cite{PhysRevA.89.033616_Queisser} in the Mott region
and it would be interesting to study their impact on the finding of this work.

As another example, higher-order correlations can mediate effective interactions
between the doublons and/or holons, giving rise to an effective Boltzmann equation
\cite{PhysRevA.100.053617,queisser2019boltzmann}.
As a consequence, the doublons and holons acquire a finite life-time whose inverse
scales with the hopping strength $T$ multiplied by the density of doublons
and/or holons \cite{queisser2019boltzmann}.
Thus, for a small density of quasi-particles, the effects of their finite
life-time can be neglected here -- unless propagation over large distances is
considered.

Going away from the small density limit considered here, one could also expect
bound states, for example excitons in the form of bound doublon-holon pairs.
The behaviour of these bound states at the interface would be another interesting
topic for future studies.

Similar arguments apply to the semiconducting region if we include a small
on-site (or Coulomb) interaction.
Assuming a small density of electron or holes in the semiconductor, the effect
of this small interaction would be sub-dominant.
For example, the electron or holes would also acquire a finite life-time whose
inverse scales with the on-site (or Coulomb) interaction strength squared,
multiplied by the density of electron or holes, and divided by the hopping
strength $T$.
As a result, the effects of a finite life-time would be even smaller than in
the Mott region.

\ack

Funded by the Deutsche Forschungsgemeinschaft (DFG, German Research Foundation) 
-- Project-ID 278162697-- SFB 1242.

\appendix
\section{Hierarchy of correlations}
\label{app_sec_hierarchy}
As mentioned in the main text, we consider reduced density matrices $\h \rho_\mu = \tr_{\cancel{\mu}}(\h \rho), \h \rho_{\mu\nu}=\tr_{\cancel{\mu}\cancel{\nu}}(\h \rho)$ etc., defined by tracing out all other lattice sites. 
These still obey the normalization condition, e.g., $\tr \h \rho_\mu=1$. Now, we follow \cite{PhysRevA.89.033616_Queisser} by splitting up the reduced density matrices into correlated and uncorrelated parts;

\bea
\label{eq:split_correlations_app}
		&\h \rho_{\mu\nu}
  = &
  \h \rho^{\rm corr}_{\mu\nu}+ \h \rho_\mu \h \rho_\nu, \nn
		&\h \rho_{\mu\nu\lambda}
  =&
  \h \rho^{\rm corr}_{\mu\nu\lambda}+\h \rho^{\rm corr}_{\mu\nu}\h \rho_\lambda +\h \rho^{\rm corr}_{\mu\lambda}\h \rho_\nu +\h \rho^{\rm corr}_{\nu\lambda}\h \rho_\mu + \h \rho_\mu \h\rho_\nu\h\rho_\lambda
  \nn
  &\h \rho_{\mu \nu \lambda \kappa}= &\h \rho_{\mu \nu \lambda \kappa}^{\rm corr}+\dots.
\ea
This is the starting point for an expansion into powers of $1/Z$ which is possible for all general lattice Hamiltonians 
\bea
	\h H = \frac{1}{Z} \sum_{\mu \nu}\h H_{\mu \nu} + \sum_\mu \h H_\mu.
\ea
In this Hamiltonian $H_{\mu \nu}$ describes two site processes, while $H_\mu$ encapsulates all one site parts like a chemical potential or Coulomb repulsion.
This includes both fermionic and bosonic systems. The time-evolution of a density operator is given by the von-Neumann equation ($\hbar = 1$)
\bea
	i \partial_t \h \rho &=& \left[ \h H,\h \rho\right] = \frac{1}{Z} \sum_{\mu \nu} \left[\h H_{\mu\nu},\h \rho\right]+\sum_\mu \left[\h H_\mu,\h \rho\right] \nn
	&=& \frac{1}{Z}\sum_{\mu \nu} \widehat{\mathcal{L}}_{\mu\nu}\h \rho + \sum_\mu \li_\mu \h \rho
\ea
where we introduced the super-operators $\li_{\mu\nu}=\big[\h H_{\mu\nu},{\h \rho}\big]$ and $\li_\mu = \big[\h H_\mu,{\h \rho}\big]$.
Inserting Eq.~\ref{eq:split_correlations_app} gives the evolution equations

\begin{equation}
     i \partial_t \h\rho_\mu 
    =
    \frac{1}{Z}\sum_{\alpha \neq\mu} \tr_\alpha \left( \li^S_{\alpha \mu} \left[ \h \rho_{\mu \alpha}^\mathrm{corr} \right]+\h \rho_\alpha \h\rho_\mu\right)+
    \li_\mu \h\rho_\mu
    \label{eq:time_evo_onsite}
\end{equation}

with the symmetrized  $\li^S_{\mu\nu}=\li_{\mu\nu}+\li_{\nu\mu}$ and
\bea
    i \partial_t \h \rho_{\mu\nu}&=& \frac{1}{Z}\sum_{\alpha \neq\mu\nu} \tr_\alpha\left(\li^S_{\alpha\mu}\h\rho_{\mu\nu\alpha}+
	\li_{\nu\alpha}^S\h\rho_{\mu\nu\alpha}\right)\nn &&+\frac{1}{Z}\li_{\mu\nu}\h\rho_{\mu\nu}+\li_\mu\h\rho_{\mu\nu}+\li_\nu\h\rho_{\mu\nu}.
	\label{eq:time_evo_twosite}
\ea
The previous equations yield the time-evolution for the two-site correlations
\bea
i \partial_t \h\rho^{\rm corr}_{\mu\nu}&=& \li_\mu \h\rho^{\rm corr}_{\mu\nu}+\frac{1}{Z}\li_{\mu\nu}\left(\h\rho^{\rm corr}_{\mu\nu}+\h\rho_\mu\h\rho_\nu\right) 
		\nn&&-\frac{\h\rho_\mu}{Z}\tr_\mu\left(\li^S_{\mu\nu}\left[\h\rho^{\rm corr}_{\mu\nu}+\h\rho_\mu\h\rho_\nu\right]\right)\nn
		&&+\frac{1}{Z}\sum_{\alpha \neq\mu\nu}\tr_\alpha\left(\li_{\mu\alpha}^S\left[\h\rho^{\rm corr}_{\mu\nu\alpha}+\h\rho^{\rm corr}_{\mu\nu}\h\rho_\alpha+\h\rho^{\rm corr}_{\nu\alpha}\h\rho_\mu\right]\right)
		\nn&&+ (\mu \leftrightarrow \nu).
\ea

We see that there is a hierarchy of equations
\bea
\label{evolution_appendix}
i\partial_t \hat\rho_\mu 
&=& 
F_1(\hat\rho_\mu,\hat\rho_{\mu\nu}^{\rm corr})
\,,\nn
i\partial_t \hat\rho_{\mu\nu}^{\rm corr} 
&=& 
F_2(\hat\rho_\mu,\hat\rho_{\mu\nu}^{\rm corr},\hat\rho_{\mu\nu\lambda}^{\rm corr})
\,,
\nn
i\partial_t \hat\rho_{\mu\nu\lambda}^{\rm corr} 
&=& 
F_3(\hat\rho_\mu,\hat\rho_{\mu\nu}^{\rm corr},\hat\rho_{\mu\nu\lambda}^{\rm corr},
\hat\rho_{\mu\nu\lambda\kappa}^{\rm corr})
\ea
and so on.
In the limit $Z\gg 1$ higher-order correlators are successively suppressed. With this, we find 
a self-consistent equation for the on-site density matrix
\begin{equation}
	i \partial_t \h\rho_\mu 
	=
	\frac{1}{Z}\sum_{\alpha \neq\mu}\tr_\alpha\left(\li_{\mu\alpha}^S\h\rho_\mu\h\rho_\alpha\right)+\li_\mu \h\rho_\mu + \mathcal{O}\left(1/Z\right).
\end{equation}
This has a solution $\h \rho_\mu^0$ that is needed for the two-site correlations
\bea
	i \partial_t \h\rho^{\rm corr}_{\mu\nu} &=& \li_\mu \h\rho^{\rm corr}_{\mu\nu} + \frac{1}{Z}\li_{\mu\nu}\h\rho^0_\mu\h\rho^0_\nu 
 -
 \frac{\h\rho^0_\mu}{Z}\tr_\mu\left(\li_{\mu\nu}^S\h\rho^0_\mu\h\rho^0_\nu\right)\nn
	&&+
 \frac{1}{Z}\sum_{\alpha \neq\mu\nu}\tr_\alpha\left(\li_{\mu\alpha}^S\left[\h\rho^{\rm corr}_{\mu\nu}\h\rho^0_\alpha+\h\rho^{\rm corr}_{\nu\alpha}\h\rho^0_\mu\right]\right)\nn
&&+
 (\mu\leftrightarrow\nu)+\mathcal{O}\left(1/Z^2\right).
\ea
In the end, the whole hierarchy is given by 
\bea
		\h \rho_\mu = \mathcal{O}\left(Z^0\right),\quad 	&&	\h \rho^{\rm corr}_{\mu\nu} = \mathcal{O}\left(1/Z\right), \nn
			\h \rho^{\rm corr}_{\mu\nu\kappa} = \mathcal{O}\left(1/Z^2\right),\quad &&	\h \rho^{\rm corr}_{\mu\nu\kappa\lambda} = \mathcal{O}\left(1/Z^3\right).
\ea

\section{Hierarchy for the extended Hubbard model}
To first order, the correlation functions are given by
\bea
\label{corr-evolution_app}
i\partial_t
\langle\hat c^\dagger_{\mu s I}\hat c_{\nu s J}\rangle^{\rm corr}
=
\frac1Z\sum_{\lambda L} T_{\mu\lambda}
\langle\hat n_{\mu\bar s}^I\rangle^0
\langle\hat c^\dagger_{\lambda s L}\hat c_{\nu s J}\rangle^{\rm corr}
\nn
-
\frac1Z\sum_{\lambda L} T_{\nu\lambda}
\langle\hat n_{\nu\bar s}^J\rangle^0
\langle\hat c^\dagger_{\mu s I}\hat c_{\lambda s L}\rangle^{\rm corr}
\nn
+
\left(U_\nu^J-U_\mu^I+V_\nu-V_\mu\right) 
\langle\hat c^\dagger_{\mu s I}\hat c_{\nu s J}\rangle^{\rm corr}
\nn
+\frac{T_{\mu\nu}}{Z}
\left(
\langle\hat n_{\mu\bar s}^I\rangle^0
\langle\hat n_{\nu s}^1\hat n_{\nu\bar s}^J\rangle^0
-
\langle\hat n_{\nu\bar s}^J\rangle^0
\langle\hat n_{\mu s}^1\hat n_{\mu\bar s}^I\rangle^0
\right) 
.
\ea
The homogeneous part, i.e. the first three lines of \ref{corr-evolution_app}, can be solved by the factorization \textit{ansatz} $\langle\hat c^\dagger_{\mu s I}\hat c_{\nu s J}\rangle^{\rm corr}
=p_{\mu s I}^* p_{\nu s J}$. Furthermore, we employ a Fourier transformation parallel to the interface, assuming a hyper-cubic lattice,

\bea
p_{\mu s I }&=\frac{1}{\sqrt{N^\parallel}}\sum_{\mathbf{k}^\parallel}p_{n,\mathbf{k}^\parallel,s I}e^{i  \mathbf{k}^\parallel \cdot\mathbf{x}_\mu^\parallel}\\
T_{\mu\nu}&=\frac{Z}{N^\parallel}\sum_{\mathbf{k}^\parallel} T_{m,n,\mathbf{k}^\parallel}e^{i \mathbf{k}^\parallel \cdot\left(\mathbf{x}_\mu^\parallel-\mathbf{x}_\nu^\parallel\right)} \,.
\eea
To simplify the notation, we drop some indices. e.g.  $p_{n,\mathbf{k}^\parallel,s I}\equiv p_\mu^I$. 
For nearest neighbour hopping the Fourier transform yields the components
\bea
T_{m,n,\mathbf{k}^\parallel}&=\frac{T^\parallel_{n,\mathbf{k}^\parallel}}{Z}\delta_{m,n}+
\frac{1}{Z}
(T^\perp_{n,n-1}\delta_{n,n-1}+T^\perp_{n,n+1}\delta_{n,n+1})\nonumber\\
T_{n,\mathbf{k}^\parallel}^\parallel&=2 T_n^\parallel \sum_{x_i}\cos(p_{x_i}^\parallel) \equiv Z T_\mathbf{k}^{\|}\,.
\eea
In the main text, this expression has been further simplified by assuming an isotropic 
hopping rate 
$T^\|_n=T^\perp_{n,n-1}=T$. The resulting effective coupled evolution equations 
are
\bea
\label{difference_app}
\left(E-U_\mu^I-V_\mu\right)p_\mu^I
+\langle\hat n_{\mu}^I\rangle^0
\sum_J T_{\bf k}^\| p_\mu^J
=
\nn
-T\frac{\langle\hat n_{\mu}^I\rangle^0}{Z}
\sum_J \left(p_{\mu-1}^J+p_{\mu+1}^J\right) 
\,.
\ea

\subsection{Semiconductor}
The semiconductor has zero Coulomb repulsion $U=0$ and a constant on-site potential $V_\mu=V$. $\hat\rho_\mu^0=\ket{\uparrow\downarrow}_\mu\!\bra{\uparrow\downarrow}$
describes the valence band and  
$\hat\rho_\mu^0=\ket{0}_\mu\!\bra{0}$ the conduction band
(at zero temperature). 
Consequently,
either one of $\langle \h n_\mu^I \rangle^0$ is zero. This yields a simple 
\bea
\left(E-V\right)p_\mu^I
+
 T_{\bf k}^\| p_\mu^J=
\frac{-T}{Z}
 \left(p_{\mu-1}^I+p_{\mu+1}^I\right) 
\,.
\ea
A propagating wave \textit{ansatz} $e^{i \kappa \mu}$ yields the well known dispersion relation $E=V-T_\mathbf{k}^\|-\frac{2T}{Z}\cos \kappa=V-T_\mathbf{k}$.
Real solutions exist for $E\in [V-T_\mathbf{k}^\|-\frac{2T}{Z},V-T_\mathbf{k}^\|+\frac{2T}{Z}]$.
\subsection{Mott insulator}
In the Mott insulator at half filling {\color{black} and without spin polarization}
\bea
\label{mean-field_app}
\hat\rho_\mu^0
=
\frac{\ket{\uparrow}_\mu\!\bra{\uparrow}+\ket{\downarrow}_\mu\!\bra{\downarrow}}{2}
\,,
\ea
the upper and lower Hubbard band solutions are related via $(E-U)p_\mu^1=E p_\mu^0$.
The plane wave \textit{ansatz} results in 
 \bea
\cos\kappa=\frac{Z}{2T}\left[\frac{E(U-E)}{E-U/2}-T_{\bf k}^\|\right]
\,.
\ea
that yields the well known dispersion relation of the Hubbard model. Real solutions for $\kappa$ exist for $E \in [E^1,E^3]$ and $E \in [E^2,E^4]$ with
\bea
&&E^{1,2}= \frac{1}{2}\left(\frac{-2T}{Z}+T_\mathbf{k}^\|+U \pm \sqrt{(\frac{-2T}{Z}+T_\mathbf{k}^\|)^2+U^2}\right) 
\nn
&&E^{3,4}= \frac{1}{2}\left(\frac{2T}{Z}+T_\mathbf{k}^\|+U \pm \sqrt{(\frac{2T}{Z}+T_\mathbf{k}^\|)^2+U^2}\right).
\nn
\ea
In the case of strong repulsion $U\gg T$ the minus or plus sign in front of the root yields the band around $E\approx 0$ and $E\approx U$, respectively.

\section{Reflection and transmission amplitudes}
\label{app:reflection_am}

In the following, we want to calculate the reflection and transmission amplitudes of the effective particle and hole solutions through the interface. Starting in the Mott {\color{black} without spin polarization}, we employ the following ansatz 
\bea
\left(
\begin{array}{c}
p_\mu^0 \\
p_\mu^1
\end{array}
\right) 
=
{\cal N} 
\left(
\begin{array}{c}
E-U \\
E 
\end{array}
\right) 
\left[e^{i\kappa\mu}+Re^{-i\kappa\mu}\right] 
\ea
and in the semiconductor we have 
\bea
\left(
\begin{array}{c}
p_\mu^0 \\
p_\mu^1
\end{array}
\right) 
=
{\cal M} 
\left(
\begin{array}{c}
x \\
1-x 
\end{array}
\right) 
{\cal{T}} e^{i\kappa\mu}
\ea
with normalization constants $\cal{N}$ and $\cal{M}$, the reflection amplitude $R$ and transmission amplitude $\cal{T}$. In the semiconductor $x=1$ yields the conduction band while $x=0$ the valance band, so $x= \langle \h n_{\mu}^0 \rangle$. Due to the doublon-holon symmetry in the semiconductor we just need to calculate the reflection for one of these cases, so in the following $x=0$. To make sense of the transmission amplitude, we need to define a probability current 
\begin{equation}
    \partial_t |p_\mu^0+p_\mu^1|^2+j_\mu-j_{\mu-1}=
iU\left[p_\mu^0 \left(p_\mu^1\right)^*-p_\mu^1 \left(p_\mu^0\right)^*\right]
\label{current}
\end{equation}
with 
\bea
\label{current_app}
j_\mu&=i\,\frac{T}{Z}\left(p_\mu^0+p_\mu^1\right)\left(p_{\mu+1}^0+p_{\mu+1}^1\right)^*+c.c. \,.
\ea
For a single mode, like done here, the ratio $p^1/p^0$ equals a real value, so the right side of Eq.~\ref{current} is 
zero. This permits the interpretation of $j_\mu$ as a probability current.
From the evolution equation \ref{difference_app} we deduct two boundary conditions, using $\mu <0$ as the Mott insulator and $\mu \geq 0$ as the semiconductor
\bea
(E-V + T_\mathbf{k}^\|) p_0^0
=
-\frac{T}{Z} \left(\frac{2E-U}{E-U}p_{-1}^0+p_1^0 \right) 
\nn 
\left(E+ \frac{T^\|_\mathbf{k}}{2}\frac{2E-U}{E-U}\right)p_{-1}^0 
=
-\frac{T}{2Z}\left( \frac{2E-U}{E-U}p_{-2}^0+p_0^0\right)
\nn
\ea
that yield the reflection amplitude $R$, Eq.~\ref{eq:refl-amplitude} as well as the transmission amplitude ${\cal T}$,   Eq.~\ref{eq:trans-amplitude}.

Using Eq.~\ref{current_app} we find the incoming, reflected and transmitted currents
\bea 
j^\mathrm{in}
&=&
2(2E-U)^2\mathcal{N}^2  \sin(\kappa_\mathrm{Mott}) ,
\nn
j^\mathrm{ref}
&=&
2(2E-U)^2\mathcal{N}^2 \sin(\kappa_\mathrm{Mott})|R|^2,
\nn
j^\mathrm{trans}
&=&
2\sin(\kappa_\mathrm{semi})|\mathcal{T}|^2.
\ea

Hence, we find the transmission probability as the ratio of the two currents
\bea
\frac{j^\mathrm{trans}}{j^\mathrm{in}}
&=&
\frac{\sin(\kappa_\mathrm{Mott})\sin(\kappa_\mathrm{semi})}{\sin(\frac{\kappa_\mathrm{Mott}+\kappa_\mathrm{semi}}{2})^2}
\ea
This quickly drops to zero for $\kappa_\mathrm{Mott} \neq \kappa_\mathrm{semi}$, as shown in Fig.~\ref{fig:trans_interface}.

\section{Tunneling through a Mott layer}
\label{app:Mott_layer}
Lastly, we want to look at the tunneling through a Mott layer sandwiched between two semiconductors. These semiconductors are both described by the same on-site potential $V$, but this can easily be generalized.
The incoming spinor is given by
\bea
\left(
\begin{array}{c}
p_\mu^0 \\
p_\mu^1
\end{array}
\right) 
=
\left(
\begin{array}{c}
x \\
1-x 
\end{array}
\right) 
\left(e^{i\kappa_\mathrm{semi}\mu} + R e^{-i\kappa_\mathrm{semi}\mu}\right),
\ea
the transmitted by 
\bea
\left(
\begin{array}{c}
p_\mu^0 \\
p_\mu^1
\end{array}
\right) 
= 
\left(
\begin{array}{c}
x \\
1-x 
\end{array}
\right) 
T e^{i\kappa_\mathrm{semi}\mu}
\ea
and inside the Mott insulator it takes the form 
\bea
\left(
\begin{array}{c}
p_\mu^0 \\
p_\mu^1
\end{array}
\right) 
=
{\cal N} 
\left(
\begin{array}{c}
E-U \\
 E
\end{array}
\right) 
\left(A e^{i\kappa\mu}+B e^{-i\kappa\mu} \right)
\ea
with the usual $\kappa_\mathrm{semi}$ and $\kappa$ in the Mott insulator. As previously, $x=0$ yields the valence band of the semiconductor, $x=1$ the conduction band. 
The layer is supposed to have a thickness of $d$ sites.  Again, we deduct boundary conditions at the interfaces and find after some algebra the transmission
\begin{equation}
\frac{j_n^\mathrm{trans}}{j_n^\mathrm{in}}=\left|\frac{e^{id \kappa }e^{-i \kappa_\mathrm{semi} d}(1-e^{i2 \kappa })(1-e^{i 2\kappa_\mathrm{semi}})}{(1- e^{i \kappa} e^{i \kappa_\mathrm{semi}})^2-e^{i 2d \kappa}(e^{i \kappa}-  e^{i \kappa_\mathrm{semi}})^2}\right|^2.
\end{equation}
This is valid for transmitted and evanescent waves and shown in for transmission through the upper Hubbard band of the Mott insulator in Fig.~\ref{fig:trans_layer}.
If both Mott bands do not match with the bands of the semiconductors, there are just decaying solutions $e^{-\kappa}$
with 
\bea
\kappa
&=&
\mathrm{arccosh}\left|\frac{Z}{2T}\left[\frac{E(U-E)}{E-U/2}-T_{\bf k}^\|\right] \right|.
\ea
There is still the possibility of tunneling.
Now, for a number of layers $d \gg 1$ and $e^{- \kappa}<1$ the transmission is given by 
\bea
\frac{j_n^\mathrm{trans}}{j_n^\mathrm{in}}\approx
4 e^{- 2d\kappa }(1-e^{-2\kappa})^2(1-\cos(\kappa_\mathrm{semi}))^2.
\ea
For $E,T \ll U$ we find the tunneling solution as
\bea
e^{- \kappa} \approx \frac{4T(E-U/2)}{Z U^2}.
\ea
For an energy exactly between the two Hubbard bands $E=U/2$ the tunneling current vanishes identically.
This was already predicted by the divergence of the Mott insulator dispersion relation in the main text.

\section{Toy Model}
\label{app:toy_model}

In order to understand the strong suppression of tunneling trough a Mott 
insulator in the middle between the upper and lower Hubbard band, 
i.e., for $E=U/2$, let us consider the following toy model consisting of 
three lattice sites in a row. 
The left and right lattice sites represent the semiconductor (where $V>0$)
and are tunnel coupled to the site in the middle which models the Mott insulator 
(with $U>0$).

This scenario is very similar to a quantum dot in the Coulomb blockade regime 
tunnel coupled to two leads.  

Now let us consider a second-order hopping process of a particle with spin 
$\uparrow$ from the right to the left site. 
If the Mott site in the middle is also occupied by a particle with spin 
$\uparrow$, the only possible tunneling sequence is 
\bea
\label{hole-channel}
\ket{0}\ket{\uparrow}\ket{\uparrow}
\stackrel{T}{\to}
\ket{\uparrow}\ket{0}\ket{\uparrow}
\stackrel{T}{\to}
\ket{\uparrow}\ket{\uparrow}\ket{0}
\,,
\ea
which corresponds to the hole channel. 
The intermediate state $\ket{\uparrow}\ket{0}\ket{\uparrow}$ has an  
energy of $2V$ which is higher than the energy $V$ of the initial and 
final states.
Thus the amplitude of this second-order hopping process is $T^\uparrow = T^2/V$,
see, e.g., \cite{splettstoesser2012}.

On the other hand, if the Mott site in the middle is occupied by a particle 
with the other spin $\downarrow$, the only possible 
(second-order) tunneling sequence is 
\bea
\label{particle-channel}
\ket{0}\ket{\downarrow}\ket{\uparrow}
\stackrel{T}{\to}
\ket{0}\ket{\downarrow\uparrow}\ket{0}
\stackrel{T}{\to}
\ket{\uparrow}\ket{\downarrow}\ket{0}
\,,
\ea
which corresponds to the particle channel. 

Now the energy of the intermediate state $\ket{0}\ket{\downarrow\uparrow}\ket{0}$
is given by $U$. 
As another difference, the $\uparrow$ particle has to pass the $\downarrow$ particle
in the middle on its way, which results in a minus sign due to the fermionic nature 
of the creation and annihilation operators.
Thus the amplitude for this second-order hopping process reads 
$T^\downarrow = -T^2/(U-V)$. 

Comparing the two results, we see that the amplitudes have opposite signs, 
$T^\uparrow = - T^\downarrow$,
in the middle between the upper and lower Hubbard band, i.e., for $V=U/2$.

If we now consider many of these rows in parallel, we may model the coherent 
propagation of a particle with transversal wavenumber ${\bf k}^\|$ from right
to left. 
Taking two rows and ${\bf k}^\|=0$ for simplicity, the combination of the two 
results~\ref{hole-channel} and \ref{particle-channel} yields 
\bea
\label{parallel}
\begin{array}{c}
\ket{0}\ket{\uparrow}\ket{\uparrow} 
\\ 
\ket{0}\ket{\downarrow}\ket{0}
\end{array}
\! + \!
\begin{array}{c}
\ket{0}\ket{\uparrow}\ket{0} 
\\ 
\ket{0}\ket{\downarrow}\ket{\uparrow}
\end{array}
\! \to 
T^\uparrow
\begin{array}{c}
\ket{\uparrow}\ket{\uparrow}\ket{0} 
\\ 
\ket{0}\ket{\downarrow}\ket{0} 
\end{array}
\! + T^\downarrow
\begin{array}{c}
\ket{0}\ket{\uparrow}\ket{0} 
\\ 
\ket{\uparrow}\ket{\downarrow}\ket{0} 
\end{array}
.
\qquad 
\ea
Due to the destructive interference between both channels,
$T^\uparrow + T^\downarrow = 0$,
the transition probability 
within the ${\bf k}^\|=0$ sector vanishes. 
This is due to the Huygens mechanism, in which the coherent propagation of a particle can be understood via the constructive interference of the amplitudes from many lattice 
sites.

Note that the same destructive interference would occur if the spin in the 
Mott insulating site is oriented in $x$-direction 
$(\ket{\uparrow}+\ket{\downarrow})/\sqrt{2}$, for example. 
Even if we consider one row only, the transport from right to left is only 
possible if that middle spin is flipped to 
$(\ket{\uparrow}-\ket{\downarrow})/\sqrt{2}$, which also destroys the coherence of the propagation.

In the true Mott state, however, each of the middle sites must be modeled by the density matrix 
$\frac{1}{2}(\ket{\uparrow}\bra{\uparrow}+\ket{\downarrow}\bra{\downarrow})$
, see Eq.~\ref{mean-field}.
Then, the cancellation in the transmission probability becomes first efficient at larger number of rows.

Since in this example the rows are not coupled with each other 
it is possible to obtain the transition probability for any number of rows $N$, including the limit of the infinite number of rows, analytically, 
based solely on the elementary amplitudes, $T^\uparrow$ and $ T^\downarrow$. 
The Mott insulating layer made up of $N$ rows with $N^\uparrow$ spin up and $N-N^\uparrow$ spin down electrons generates a tunnelling probability
\begin{equation}
    P_{N^{\uparrow}} = \frac{|N^\uparrow T^\uparrow + (N-N^\uparrow) T^\downarrow|^2}{N^2}.
\end{equation}

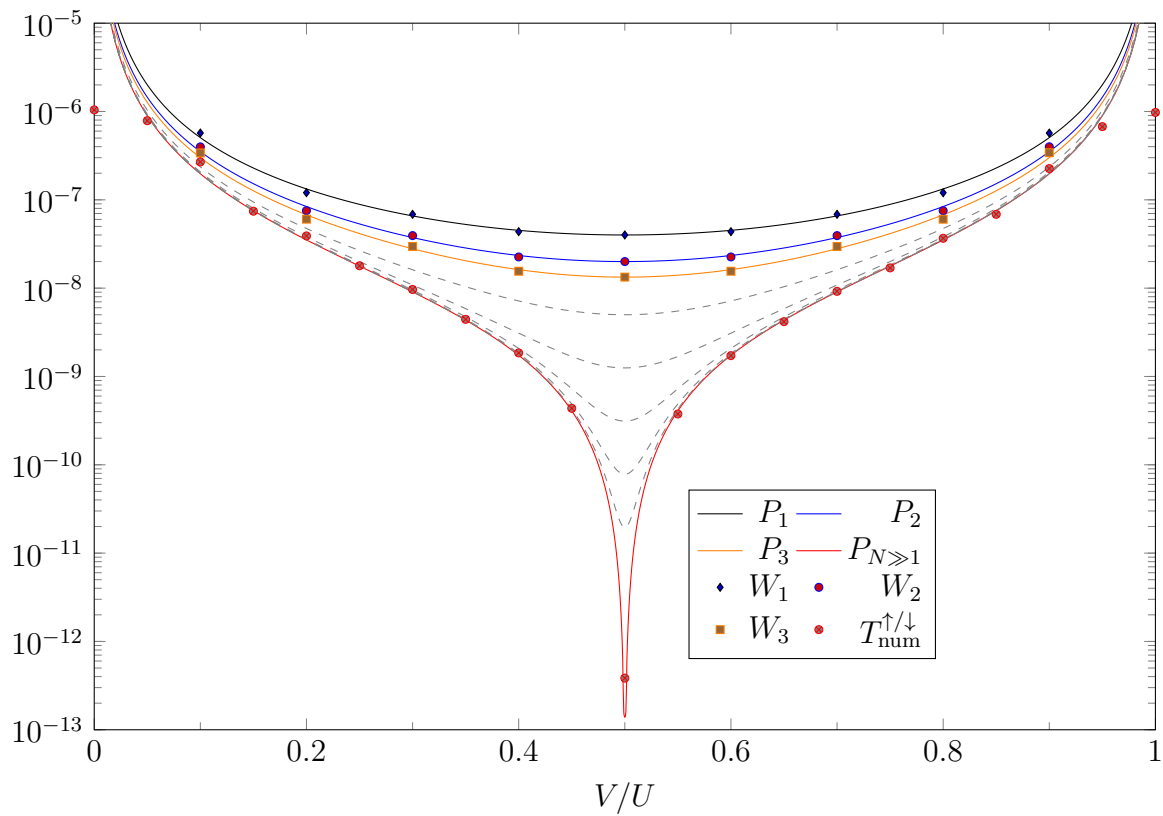
\begin{figure}[t] 
    \centering
   	\begin{tikzpicture}
		\begin{axis}[
			xmin = 0., xmax = 1,
            ymin=1e-13,
            ymax=1e-5,
			xtick distance = .2,
    max space between ticks=20,
			minor tick num = 1,
			major grid style = {lightgray},
			minor grid style = {lightgray!25},
			width = \linewidth,
			height = 0.7\linewidth,
			xlabel = {$V/U$},
			ylabel = {},
			legend cell align = {right},
            legend style={at={(0.56,.22)},anchor=west},
            ymode = log,
            legend columns=2
			]
		
            \addplot[smooth, black] table[x index=0, y index=1] {pics/alles_pert.dat};
            \addplot[smooth, blue] table[x index=0, y index=2] {pics/alles_pert.dat};
            \addplot[smooth, orange] table[x index=0, y index=3] {pics/alles_pert.dat};
            \addplot[smooth, red] table[x index=0, y index=4] {pics/alles_pert.dat};
    
            \addplot+[ only marks, black, mark size=1.5pt ] table[x index=0, y index=1] {pics/alles_num.dat};
            \addplot+[ only marks, blue, mark size=1.5pt ] table[x index=0, y index=2] {pics/alles_num.dat};
            \addplot+[ only marks, orange, mark size=1.5pt ] table[x index=0, y index=3] {pics/alles_num.dat};
            \addplot+[ only marks, red, mark size=1.5pt ] table[x index=0, y index=1] {pics/alles_inftab_num.dat};

             \addplot[thin, smooth, gray, dashed] table[x index=0, y index=1] {pics/alles_pert_grosseN.dat};
             \addplot[thin, smooth, gray, dashed] table[x index=0, y index=2] {pics/alles_pert_grosseN.dat};
             \addplot[thin, smooth, gray, dashed] table[x index=0, y index=3] {pics/alles_pert_grosseN.dat};
             \addplot[thin, smooth, gray, dashed] table[x index=0, y index=4] {pics/alles_pert_grosseN.dat};
             \addplot[thin, smooth, gray, dashed] table[x index=0, y index=5] {pics/alles_pert_grosseN.dat};
			\legend{$P_1$,$P_2$,$P_3$,$P_{N\gg 1}$,$W_1$,$W_2$,$W_3$,$T^{\uparrow/\downarrow}_\mathrm{num}$}
						
		\end{axis}

	\end{tikzpicture}
    \caption{(Color online) Transition probabilities $P_N$ for coherent waves in the toy model for different numbers of rows $N$. The lines give the perturbation theory result, the black, blue and orange markers depict the properly rescaled numerical expectation values of the wave behind the Mott sites $W_N$. For $N=\infty$ the red circles give the probability using numerically obtained amplitudes $T_\uparrow$ and $T_\downarrow$ in Eq.~\ref{eq_pert_theo_gen}. The gray dashed lines give $P_N$ for $N=2^j$ with $j=3,5,7,9,11$. The hopping strength is $T=0.01\, U$.}
    \label{fig:vgl_pert}
\end{figure}

Summing up over all spins configurations, $N^\uparrow$, with the appropriate combinatorial factors yields the probability
\begin{equation}
P_N= \frac{N+1}{4 N} \left( |T^\uparrow|^2 + |T^\downarrow|^2 \right) + \frac{N-1}{2 N}\mathrm{Re}(T^\uparrow T^{\downarrow *}).    
\end{equation}

This probability has a minimum for $T^\uparrow = - T^\downarrow$, which due to symmetry occurs in the middle of the gap for $V=U/2$, and falls as $1/N$.
In the limit $N\gg 1$ we find 
\begin{equation}
\label{eq_pert_theo_gen}
    P_{N \gg 1} = \frac{1}{4}\, |T^\uparrow+T^{\downarrow}|^2
\end{equation}
which vanishes in the middle of the gap.
This destructive interference is similar to the suppression of tunneling 
by a strong disorder potential. 
Using the perturbation theory amplitudes $T^\uparrow, T^\downarrow$  given above we find for $N\gg 1$
\begin{equation}
\label{pert_theo}
    P_{N\gg 1} = T^4 \frac{(V - \frac{U}{2})^2}{V^2 (U-V)^2}
\end{equation}
which is zero exactly at $V=U/2$.

The perturbative results have been compared with those obtained numerically. For a small number of rows $N$, the comparison of $P_N$ is done with numerically obtained expectation values of the plane wave behind the Mott layers $W_N$. The initial state, composed of the Mott sites and the plane wave in the left semiconductor layer, is evolved forward in time and the expectation value of this wave behind the Mott region is measured. The $W_N$ are rescaled in order to match the transmission probabilites at $V=U/2$ (since their normalization is unknown). For $N=1, \,2,\,3$ Fig.~\ref{fig:vgl_pert} shows the $P_N$ as functions of $V/U$ (solid lines) together with the numerical results (markers with the same color). The numerical results agree very well with the perturbation theory (away of the poles at $V=0$ and $V=U$): The $1/N$ behaviour in the middle of the gap and the shapes of the curves are confirmed. Furthermore, we calculated numerically the amplitudes $T^\uparrow$ and $T^\downarrow$ for a single row 
from the probabilities $\left|\langle \uparrow\,\, \uparrow 0 |U(t) |0 \uparrow\,\, \uparrow \rangle\right|^2$ and $\left|\langle \uparrow\,\, \downarrow0 |U(t) |0 \uparrow \,\,\downarrow \rangle\right|^2$, respectively, using the full Hamiltonian. For short times, these probabilities grow quadratically in time with the prefactor equal to the second order hopping process \cite{PhysRevResearch.4.L032045}.
By inserting them into Eq.~\ref{eq_pert_theo_gen} we obtain the red squares in Fig.~\ref{fig:vgl_pert}, which also agree very well with $P_{N\gg 1}$, including the drop in the transmission probability at $V= U/2$. The development of this suppression with growing $N$ is indicated by the gray dashed lines.

\vfill
\bibliographystyle{iopart-num}
\bibliography{paper_bib}

\end{document}